\documentclass{article}

\usepackage{PRIMEarxiv}
\usepackage{placeins}
\usepackage{booktabs}
\usepackage{multicol}
\usepackage[numbers]{natbib}
\usepackage{array}
\usepackage[utf8]{inputenc} % allow utf-8 input
\usepackage[T1]{fontenc}    % use 8-bit T1 fonts
\usepackage{hyperref}       % hyperlinks
\usepackage{url}            % simple URL typesetting
\usepackage{booktabs}       % professional-quality tables
\usepackage{amsfonts}       % blackboard math symbols
\usepackage{nicefrac}       % compact symbols for 1/2, etc.
\usepackage{microtype}      % microtypography
\usepackage{lipsum}
\usepackage{fancyhdr}       % header
\usepackage{graphicx}       % graphics
\usepackage[textsize=small]{todonotes}
\usepackage[switch]{lineno}
\usepackage{url}
\usepackage{amsmath}
\usepackage{mathtools, nccmath, textcomp}
\graphicspath{{media/}}     % organize your images and other figures under media/ folder

\title{Representation Learning for Appliance Recognition: \\ A Comparison to Classical Machine Learning
}

\author{
  Matthias Kahl, Daniel Jorde, Hans-Arno Jacobsen \\
  Technical University\\
  Munich \\
  \texttt{\{matthias.kahl, daniel.jorde}@tum.de\} \\
}

\begin{document}
\maketitle

\begin{abstract}
Non-intrusive load monitoring (NILM) aims at energy consumption and appliance state information retrieval from aggregated consumption measurements, with the help of signal processing and machine learning algorithms. Representation learning with deep neural networks is successfully applied to several related disciplines. The main advantage of representation learning lies in replacing an expert-driven, hand-crafted feature extraction with hierarchical learning from many representations in raw data format. In this paper, we show how the NILM processing-chain can be improved, reduced in complexity and alternatively designed with recent deep learning algorithms. On the basis of an event-based appliance recognition approach, we evaluate seven different classification models: a classical machine learning approach that is based on a hand-crafted feature extraction, three different deep neural network architectures for automated feature extraction on raw waveform data, as well as three baseline approaches for raw data processing. We evaluate all approaches on two large-scale energy consumption datasets with more than 50,000 events of 44 appliances. We show that with the use of deep learning, we are able to reach and surpass the performance of the state-of-the-art classical machine learning approach for appliance recognition with an F-Score of 0.75 and 0.86 compared to 0.69 and 0.87 of the classical approach.
\end{abstract}

% keywords can be removed

\section{Introduction} \label{sec:Introduction}
%\IEEEPARstart{T}{he} vast amount of climate scientists agrees to the human influenced climate change. More than 65\% of the worldwide electrical energy is produced by non-carbon-neutral fossil fuels \cite{Agency2018} which releases the greenhouse gas $CO_2$ into the atmosphere. A higher carbon concentration causes temperature increase of the atmosphere \cite{Hansen1981}.
%\IEEEPARstart{A}{round}
Around 27\% of the worldwide electrical energy consumption goes to the residential sector \cite{Agency2018}. Residential consumers are not aware of the energy consumption of the individual appliances due to insufficient data provided in their energy billing report. One way to retrieve disaggregated consumption data is the per-appliance metering on each appliance of interest. The downsides of the so-called intrusive load monitoring, are high costs and increased energy consumption from multiple measurement units and a significant intervention into the electric circuit. The main goal of non-intrusive appliance monitoring (NILM) lies in providing detailed energy consumption feedback on an appliance-level with minimal to no intervention into the electrical grid to increase the acceptance for electricity monitoring systems. NILM measures only at the aggregated central point (electric cabinet) with a single \emph{intelligent} meter. This way, installation effort, hardware cost and energy consumption of the metering system itself are significantly reduced. State-of-the-art machine learning and signal processing techniques enable the retrieval of consumption information on an appliance-level (load disaggregation). Use cases for NILM systems include the support for building automation systems \cite{Norford1996}, elderly care \cite{Alcala2017}, demand response \cite{Lin2015} and predictive maintenance of hardware and machinery \cite{Armel2013}.

%Many NILM-focused studies have been published since the late 1980s. The entrance of smart meters into the electricity market brought a new wave of publications in the late 2000s from a still growing and active community \cite{Kelly2016}.

The still remaining error margin of these systems is a major drawback that prevents NILM systems from being rolled out in the form of commercial products. Appliance event detection and classification are challenging tasks that significantly influence the overall NILM performance and acceptance by the consumers. Deep neural networks are a leading technology in speech and visual object recognition. To gain insights on the potential of representation learning for appliance recognition, we implemented multiple deep learning algorithms and conducted a broad comparison to several classical machine learning approaches on raw waveform data for appliance recognition. The evaluation includes an expert-aided, 212-dimensional hand-crafted feature extraction model, three baseline raw data processing models, four different classifiers and three deep neural network architectures with their network parameter configuration for household consumption data. In summary, our contributions are as follows
\begin{enumerate}
  \item Design of an end-to-end representation learning approach for appliance recognition.
  \item Comparison of seven appliance recognition models.
  \item Detailed presentation of the deep neural network parameter configuration for appliance recognition.
\end{enumerate}

%An additional outcome of this paper is the introduction of a representation learning approach that outperforms all introduced models.

The remainder of this paper is organized as follows: In Section\,\ref{sec:Background}, we introduce NILM and deep learning. Section\,\ref{sec:Related_Work} highlights recent studies on appliance recognition and deep learning approaches. In Section\,\ref{sec:Architecture}, we describe our appliance recognition architecture. The different recognition models are explained in Section\,\ref{sec:Experiments}. In Section\,\ref{sec:Results}, we present results, before concluding in Section\,\ref{sec:Conclusions}.

%\hfill mds
%\hfill August 26, 2015

\section{Background} \label{sec:Background}
We begin with an introduction into NILM and explain deep learning in general and the employed architectures and performance metrics in particular.
\subsubsection*{NILM} \citet{Hart1992} introduced NILM as a technique to investigate energy consumption with appliance specific characteristics. \emph{Active power}, \emph{reactive power} and \emph{apparent power} have become a standard set of observational variables for NILM purposes. Since then, many other characteristics have been developed as features for classical machine learning algorithms to distinguish electrical appliances. These features can be divided into spectral and temporal domain such as \emph{harmonics}, \emph{total harmonic distortion} and \emph{spectral flatness}, as well as \emph{crest factor}, \emph{form factor} and \emph{VI-trajectory}. The paper of \citet{Sadeghianpourhamami2017} and our previous work \cite{Kahl2017} give an overview and evaluation of many features for appliance classification purposes.

\subsubsection*{Deep Learning}
Deep learning is a specialization of representation learning and a generic term for learning strategies for computational models, built from multiple processing layers that allow learning from representations of data in different levels of abstraction. Deep neural networks are currently the state of the art in speech and visual object recognition \cite{LeCun2015, Hinton2012}. The main difference to classical machine learning lies in the automated finding of discriminative features for classification. Domain-specific expert knowledge for feature extraction is not necessary \cite{LeCun2015}. Usually, a large portion of data in raw images, raw audio as well as raw energy consumption signals can be ignored due to redundancy and irrelevance regarding the discriminative potential. To retrieve the essence of the raw data, we select three deep neural network architectures.

\paragraph{Deep Autoencoder (AE)}
AEs are feedforward nets with a different number of neurons in the inner coding layer. The goal of an AE is to reach the same output as the input by propagating the input through the different dimensional coding layer in the middle. In our case, the AE's target is to reduce the number of representative neurons in the inner coding layer. The AE is built of an encoding section in which the input data is reduced and a decoding section in which the reduced codings are upscaled to reproduce the input. The output of the encoding section can be seen as a lower dimensional representation of the input \cite{Geron2017} that went through a bottleneck, keeping only the essence of the data.

\paragraph{Deep Convolutional Neural Network (CNN)}
CNNs are designed to process signals that follow the principle of locality. Natural signals can be efficiently processed by local connections, shared weights, pooling and the use of multiple layers \cite{LeCun2015}. CNNs benefit from the typical hierarchical composition of natural signals.

\paragraph{(Deep) Convolutional Autoencoder (CAE)}
CAEs follow the same basic architecture as a standard AE. The hidden layers are replaced by convolutional layers, inheriting the advantages of locality, shared weights and pooling.

\subsubsection*{Performance Metrics}
The metrics are calculated using the unweighted macro-average of all class-wise results. To evaluate the classification performance for each class in a multi-class problem, Precision (PR), Recall (RE) and F-Score are common metrics and implemented using the class-wise True Positives (TP), False Positives (FP) and False Negatives (FN) as follows:

{\small%
\begin{align*}
\textbf{F-Score} &= 2\cdot\mfrac{\text{PR}\cdot{\text{RE}}}{\text{PR + RE}} &~
\textbf{PR} &= \mfrac{\text{TP}}{\text{TP + FP}} ~&
\textbf{RE} &= \mfrac{\text{TP}}{\text{TP + FN}} \\
\end{align*}%
}%
\vspace{-5mm}

\section{Related work} \label{sec:Related_Work}
\begin{figure}[htbp]
\centering
\includegraphics[width=0.6\linewidth]{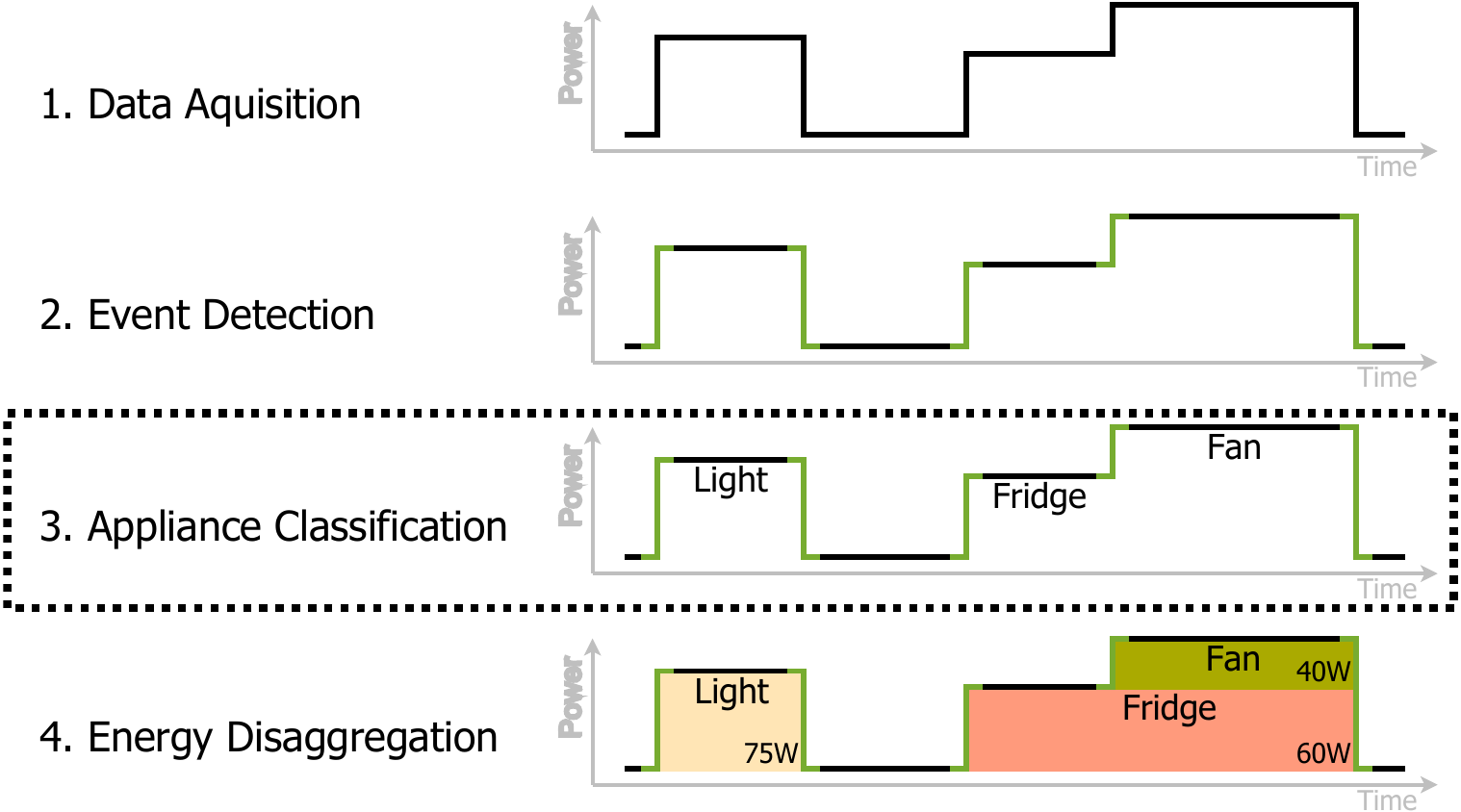}
\caption{The focus of this work is on an appliance classification of the general NILM processing-chain. The actual appliance switch-on and switch-off events are retrieved with the help of the provided metadata and low-frequency measurements of the datasets and therefore considered as known in advance.}
\label{fig:nilmsteps}
\end{figure}

Energy consumption feedback with NILM can be generally divided into 4 steps \cite{Anderson2012a}: (1) data acquisition, (2) event detection, (3) appliance classification, and (4) energy disaggregation, see Figure~\ref{fig:nilmsteps}. Each NILM step is a relevant subject of research with individual approaches and studies \cite{Haq2018,Wild2015,Hassan2014,Wichakool2015}. The focus of this paper lies in the NILM subtask (3): appliance classification, which can be regarded as a multivariate classification problem.

%The focus in this work lies in the NILM subtask appliance recognition which can be modeled as a multivariate classification problem.

The NILM community evaluated several classifiers in recent years. Hidden Markov Models are used in the work of \citet{Kolter2012} and \citet{Zhong2014} for appliance disaggregation, while \citet{Kramer2015} and \citet{Du2012} focus on K-Nearest Neighbor and Support Vector Machines for appliance recognition. NILM studies either belong to the low-frequency or high-frequency domain. Approaches that work on measurements sampled at less than 1\,Hz are usually considered low-frequency while measurements with a sampling frequency more than twice as high as the mains frequency are usually characterizing the waveform and considered as high-frequency. Studies that evaluate approaches in the low-frequency domain often aim for solutions of actual energy provider driven smart meters, since their sampling frequency is usually limited due to privacy concerns. High-frequency sampled waveform measurements usually aim for in-house monitoring solutions, driven by the consumer. \citet{Armel2013} shows that an increase in sampling frequency also causes an increase in the number of appliances that can be distinguished.

Since waveform-based appliance energy consumption shares similarities to audio signals (signal envelope and appliance events), audio features can be successfully applied for appliance recognition \cite{Kahl2017} which motivates further studies on deep neural networks for appliance recognition. In computer vision, deep CNNs received a lot more attention due to the paper of \citet{Krizhevsky2012}, who reduced the error rate for visual object recognition by almost half. The publication of \citet{Hinton2012} shows the performance improvements of deep neural networks in speech recognition from four renowned research groups. CNNs can be successfully applied to visual and audio related classification problems.

\citet{Lee2017} propose an approach, based on a deep CNN for music tagging on sample-level (raw data). The results of the 10+ layer-sized deep neural networks are comparable to the previous state-of-the-art performances. \citet{Dai2017} use very deep CNNs (up to 34 layers) to classify environmental sounds on raw data. The best architecture comprises 18 hidden layers and reaches the performance of a CNN with the audio spectrogram as input. The complex net architecture of the two approaches shows the potential for promising results on the one side, and that working on raw waveform data is challenging on the other side, especially in finding the right net architecture. \citet{Jorde2018} propose the first approach on an appliance classification that uses deep neural networks on raw measurements. To overcome the issue of a small training-set, data augmentation and a one-against-all classifier composition were implemented to reach state-of-the-art classification performances.

Our approach considers the evaluation of three deep learning architectures (AE, CAE, CNN) in comparison with a comprehensive classical machine learning approach that uses 36 hand-crafted features. The AE and CAE are used for automated feature extraction from raw data, while the CNN is implemented as an end-to-end classification system to gain the full potential of deep learning architectures. The goal is to design an appliance recognition system that keeps the amount of preprocessing and domain-specific knowledge for feature extraction to a minimum, still reaching state-of-the-art classification performances.

% Our contribution:
% - machine vs. representation learning benchmark
% - no data augmentation
% - all-in-one classifier
% - real household datasets

% OK general appliance classification approaches
% OK works that show DL can reach better Results
% ok works that use DL and are similar to appl. recognition
% ok explain Daniels work

% !TEX root = start_file.tex

\section{The Appliance Recognition Process} \label{sec:Architecture}
\begin{figure}[htbp]
\centering
\includegraphics[width=0.7\linewidth]{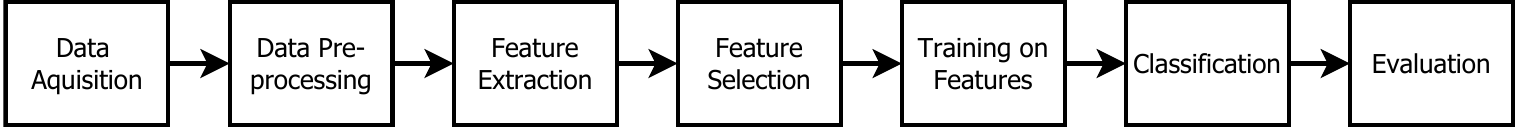}
\caption{The architecture of a classical machine learning approach always requires domain specific expert knowledge for finding types of features with class-discriminative potential in the hand-crafted feature extraction process step three.}
\label{fig:mlArchitecture}
\end{figure}

\begin{figure}[htbp]
\centering
\includegraphics[width=0.7\linewidth]{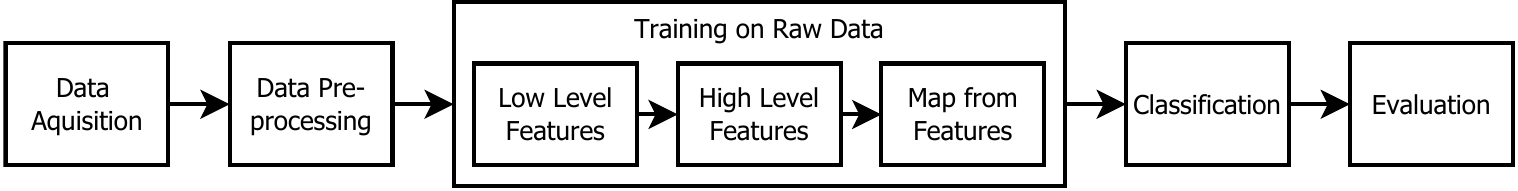}
\caption{The architecture of a representation learning approach tries to replace the hand-crafted feature extraction with automated, hierarchical feature extraction without the need of domain specific expert knowledge.}
\label{fig:rlArchitecture}
\end{figure}

We implemented two different appliance recognition systems, a classical machine learning, and a representation learning approach. The typical architectures of these learning systems can be seen in Figure~\ref{fig:mlArchitecture} and \ref{fig:rlArchitecture}. We chose two publicly available energy consumption datasets of a residential and office environment. The datasets are the most suitable selection of the publicly available datasets for our experiments on the selected deep learning algorithms due to their considerably different set of appliances and usage patterns. Since we use existing datasets, data acquisition does not play any role in this work. Further acquisition details regarding the datasets can be found in the work of \citet{Kelly2015} (UK-DALE) and \citet{Kriechbaumer2018} (BLOND-50).

\subsubsection*{UK-DALE}
The UK Domestic Appliance-Level Electricity (UK-DALE) dataset consists of more than 4 years of energy consumption measurements for a residential building (house-1) with a high number of appliances of many different types. For our experiments, we considered measurements from 2013-04-22 to 2015-01-05. The dataset comprises low-frequency, non-equidistant sampled smart plug measurements (\o ~\texttildelow 1/6\,Hz) for each observed appliance (per-appliance signals) and high-frequency sampled measurements (16\,kHz) from a custom sound card meter at the electric cabinet (aggregated signal). The per-appliance measurements allow a coarse determination of appliance events and power consumption to extract the relevant segments from the aggregated signal.

\subsubsection*{BLOND-50}
The Building-Level Office eNvironment Dataset (BLOND) comprises energy consumption measurements from an office building with a high number of appliances of only a few different types. This appliance and appliance type distribution is the main difference between the datasets, covering a wide spectrum of real environments. The BLOND-50 subset comprises 213 days of recording with 50\,kHz sampling frequency for the aggregated signal at the electric cabinet and 90 individually observed sockets for the per-appliance measurements with 6.4\,kHz sampling frequency.

Our appliance recognition process uses only the first 500\,ms of the appliance startup current and voltage as the baseline for the hand-crafted and automated feature extraction of the considered algorithms, categorizing it as a so-called event-based approach.


\subsection{Data Preprocessing and Event Detection}
The appliance event time-stamps can be approximated with the help of the additional per-appliance measurements, which are provided in both datasets (1/6\,Hz for UK-DALE, 6.4\,kHz for BLOND-50). A simple threshold-based event detection algorithm with appliance individual thresholds is used to detect events in these per-appliance measurements (see Table~\ref{tab:eventThresholds}). The resulting appliance event time-stamps are used to extract segments from the high frequency aggregated signal. The power-related switch-on threshold $\Delta_\uparrow$ in Watts defines the power state at which an appliance is considered switched on - similarly for the switch-off threshold $\Delta_\downarrow$.

\begin{table}[htbp]
\centering
\small
\caption{UK-DALE Appliance Event Thresholds and Quantity}
\label{tab:eventThresholds}
\setlength{\tabcolsep}{1.75mm}
\begin{tabular}{lrrrclrrr}
\textsc{Appliance} & $\Delta_\uparrow$ & $\Delta_\downarrow$ & \# Events & & \textsc{Appliance} & $\Delta_\uparrow$ & $\Delta_\downarrow$ & \# Events \\ \midrule
Boiler & 70 & 20 & 1,701 & & LCD Office & 30 & 4 & 1,337 \\
Solar Thermal Pump & 40 & 20 & 5,221 & & Breadmaker & 400 & 20 & 649 \\
Laptop & 20 & 2 & 498 & & Amp Livingroom & 18 & 10 & 945 \\
Washing Machine & 1,500 & 1 & 506 & & Hoover & 400 & 10 & 392 \\
Dishwasher & 100 & 20 & 885 & & Coffee Machine & 1,000 & 10 & 38 \\
TV & 70 & 10 & 907 & & Hair Dryer & 100 & 20 & 713 \\
Kitchen Lights & 70 & 20 & 4,765 & & Straightener & 300 & 5 & 264 \\
HTPC & 70 & 20 & 1,169 & & Iron & 1,000 & 10 & 147 \\
Kettle & 2,000 & 10 & 2,674 & & Gas Oven & 35 & 10 & 492 \\
Toaster & 1,000 & 10 & 1,495 & & Office Fan & 20 & 2 & 78 \\
Fridge & 70 & 10 & 15,766 & & LED Printer & 800 & 3 & 159 \\
Microwave & 500 & 10 & 3,363 & & & & & \\ \bottomrule
\end{tabular}
\end{table}

The UK-DALE dataset comprises 52 appliances of several types. For this work, we use a subset of 23 appliances by selecting only one appliance per appliance type and ignoring low-power devices such as \emph{ADSL Router}, \emph{Ipad Charger} and \emph{Baby Monitor} that are potentially undetectable in a noisier aggregated signal, due to their low power consumption. The sample numbers of the remaining appliance classes are very heterogeneous and vary between 38 for the coffee machine and 15,766 for the fridge, due to their natural consumer pattern (see Figure~\ref{tab:eventThresholds}).
%that are unlikely of being detectable at all in a noisy aggregated signal, due to their small power impact relative to the noise intensity.

For the BLOND-50 dataset, a general power threshold of 25\,W defines a switch-on and switch-off event. All 21 occurring appliances of the chosen time span are included in the remaining appliance types: \emph{Laptop} (13), \emph{Monitor} (5), \emph{PC} (2) and \emph{Printer} (1), amounting to 9,321 appliance samples.

Regarding UK-DALE, with the known appliance event time-stamp from the per-appliance measurements, the high-frequency aggregated measurements are observed in a 20\,s time-window for the exact event position. For each appliance event that is found in the per-appliance measurements, a segment of the first 500\,ms is extracted from the high-frequency aggregated measurements at the corresponding time-stamp. These 500\,ms long calibrated startup-transients are the baselines for all following considerations.

It is important to keep in mind that there are two sources of event inaccuracies in this step for UK-DALE. The first lies in the event detection on the low-frequency per-appliance measurements. With 1/6\,Hz, the sampling rate is too low to detect short-term consumption patterns. The second lies in finding the exact event position in the 20s time-window in case of multiple occurring events in that time window. The probability to extract the correct appliance segments equals the reciprocal of the number of individual appliance events in the 20s time-window. Reliable error estimation is unfortunately not possible. Since the per-appliance measurements are sampled with 6.4\,kHz, the event time-stamps are accurate enough to not cause these issues for BLOND-50.

%Further investigation on that issue reveals that in 69\,\% of the cases, the temporal chronological distance between 2 events is lower than 20\,s. Considering the fact that in around 21\,\% of these cases the 2 events are of the same appliance and only in 50\,\% of these cases the wrong event is captured, the theoretical appliance event mismatch rate would lie in a range of 27\,\%. To overcome that issue, we reuse the appliance-specific switch-on threshold $\Delta_\uparrow$ to find the actual event transient in the aggregated high-frequency signal.

%\todo{maybe add numbers}

%The main architectural differences lie in the third of the five typical steps of learning systems. While

\subsection{Hand-Crafted Feature Extraction}
%The feature extraction can be seen as the most sophisticated part of a machine learning pipeline.
For this work, we extracted 36 features that are introduced and explained in our previous work \cite{Kahl2017}. These features comprise traditional electricity metrics such as \emph{active \& reactive power}, \emph{admittance}, \emph{crest factor} and \emph{phase shift}, audio processing features such as \emph{harmonics}, \emph{wavelet analysis} and \emph{total harmonic distortion}, a selection of MPEG7 audio descriptors\footnote{Since high frequency sampled energy consumption shares similarities to audio and music data, the MPEG7 audio descriptors contain features with significant discriminative potential for electrical appliance} and novel metrics such as \emph{max-inrush-ratio} and \emph{inrush-current-ratio}. The chosen features are one- and multidimensional with a total number of 212 dimensions. With these features, we reach very high appliance classification performances (F-Score between 0.76 and 1.0) across household-focused subsets of the four publicly available datasets WHITED~\cite{Kahl2016}, PLAID~\cite{Gao2014}, UK-DALE~\cite{Kelly2015} and BLUED~\cite{Anderson2012} with the standard classifier (KNN, SVM, LDA, BDT).

%With these features - a subset of these features, respectively - it is possible to reach high appliance classification performances across household-focused subsets of the four publicly available datasets WHITED\cite{Kahl2016} (F-Score: 1.0), PLAID\cite{Gao2014} (F-Score: 0.96), UK-DALE\cite{Kelly2015} (F-Score: 0.79) and BLUED\cite{Anderson2012} (F-Score: 0.76).

\subsection{Autoencoder}
We implemented three different AE architectures that are designed to reduce the raw waveform data to a 212-dimensional feature space. The set comprises a one, two and three-layered encoding and decoding architecture with different dimensionality. The AEs have a mirror-like design, which means that the decoding layers are identical to the encoding layers, but in reverse order.

%The one layered design reduces the raw data directly from 8000 / 25000 (UK-DALE / BLOND) dimensions to 212. The two-layered design reduces to 1800 / 2000 (UK-DALE / BLOND) in the first layer and the three-layered reduces by factor 2, 4 and 5

\subsection{Convolutional Neural Network Architecture}
To ensure an automated hierarchical feature extraction with the 1-dimensional CNN, we applied a sampling and mains frequency ($f_s, f_0$) dependent layer architecture. The goal is to reduce the layer inputs with max-pooling layers in a way that the output dimension of the last convolutional layer is identical to the number of mains cycles ($n_p=25$ for 50\,Hz mains frequency) of the 500\,ms segments. The number of convolutional layers ($n_l$) and its kernel sizes ($k$) are calculated as follows:

{\small
$$
\vec{k}=sort_{desc}(prime\_factorization(\frac{f_s}{f_0})), \hspace{0.5cm}
n_l=\#k
%8000 / 25 = 320, \hspace{0.5cm} 320 / 5 / 2 / 2 / 2 / 2 / 2 = 1
$$
}

\begin{figure}[htbp]
\centering
\includegraphics[width=0.9\linewidth]{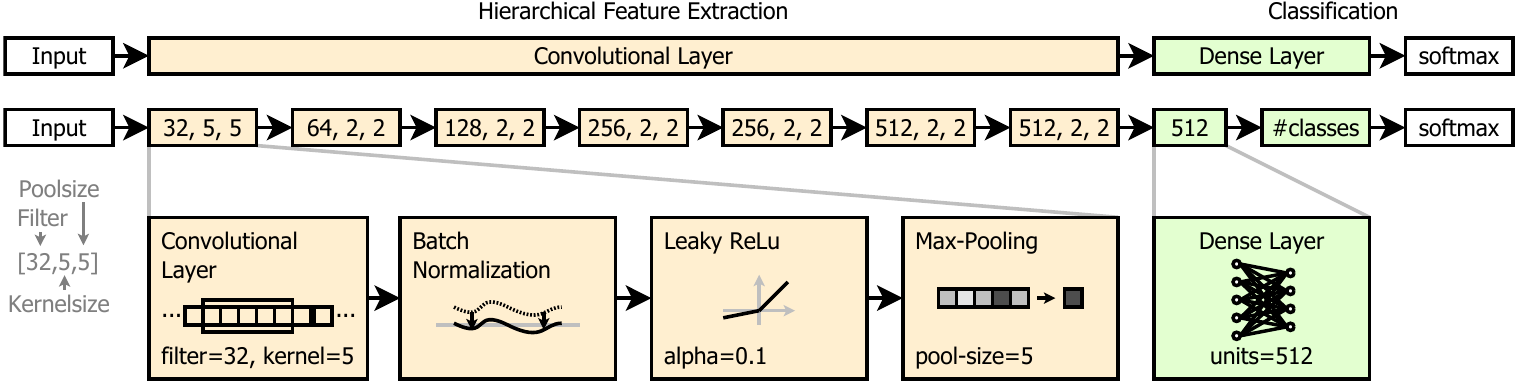}
\caption{The architecture of the end-to-end implementation of the CNN for the UK-DALE dataset.}
\label{fig:ConvArchitecture}
\end{figure}

Therefore, the pool-sizes of the max-pooling layers for the UK-DALE dataset are $[5,2,2,2,2,2,2]$ (see Figure~\ref{fig:ConvArchitecture}) while they are $[5,5,5,2,2,2]$ for the BLOND-50 dataset.

\subsection{Convolutional Autoencoder}
\begin{figure}[htbp]
\centering
\includegraphics[width=0.9\linewidth]{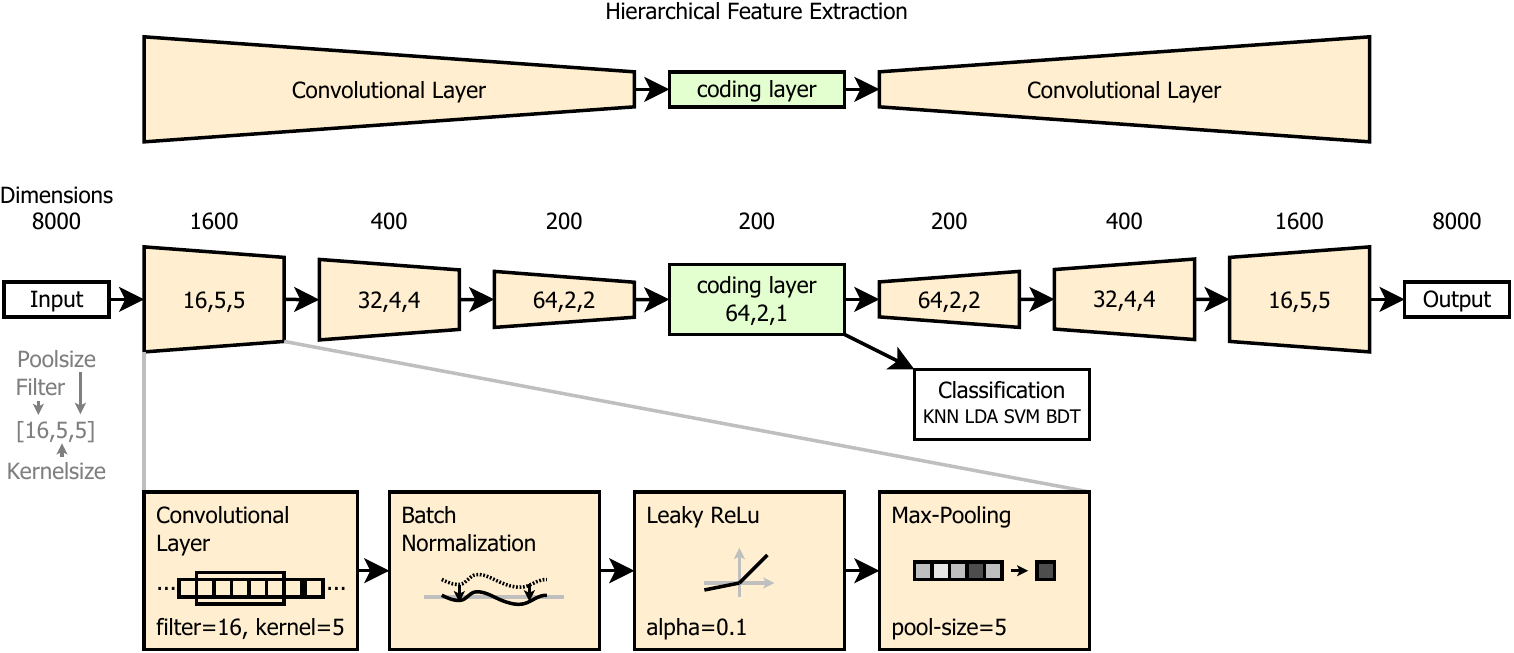}
\caption{The architecture of the CAE network for the UK-DALE dataset.}
\label{fig:CAEArchitecture}
\end{figure}

The advantage of CAEs lies in the use of convolutional layers inside an AE network. They benefit from a better locality of natural signals while general AEs handle each input dimension as global. In other words, the order of the input dimensions does not matter for a general, fully connected AE as opposed to using convolutional layers. To keep the model trainable, we implemented three encoding, one coding, and three decoding layers. To ensure comparability with the hand-crafted features extraction approach, we reduced the feature dimensions to 200. This is the closest we could reduce 8,000 and 25,000 to the 212 dimensions of the hand-crafted features approach, using only integer divisors. Therefore, the encoder and decoder pool-sizes (divisors) of the max-pooling layers are $[5,4,2]$ and $[2,4,5]$ for UK-DALE (see Figure~\ref{fig:CAEArchitecture}), while they are $[5,5,5]$ and $[5,5,5]$ for BLOND-50. The number of filter and the kernel sizes of each layer are identical for both datasets, see Figure~\ref{fig:CAEArchitecture}.

\subsection{Feature Space Transformation} \label{sec:featSpaceTransform}
We implemented two different ways of feature space transformation to avoid undesired feature weighting, caused by different value ranges across the dimensions: the variance-normalization $x_{var}'$ and max-normalization $x_{max}'$, calculated from $x$ as unprocessed feature vector:

{\small
$$
x_{var}' = \frac{x - mean(x)}{var(x)}, \hspace{1.25cm}
x_{max}' = \frac{x}{abs(x)}
$$
}

For the variance normalization, $mean(x)$ and $var(x)$ are calculated from the training-set and test-set independently.

% !TEX root = start_file.tex

\section{Experiments} \label{sec:Experiments}
%Our experimental setup that represents a machine learning approach is implemented using 36 features from different research domains.
As a preparation step, all samples of both datasets are shuffled and stratified split into 80\,\% training samples and 20\,\% test samples to ensure the same class sample number heterogeneity in training and test-set. To better assess the results of the machine and representation learning approaches, we additionally implemented 3 classification models as baseline-reference models. In total, 7 different models are evaluated. The classification is implemented with four common classifiers: K-Nearest Neighbor (KNN \cite{Runkler2012}), Linear Discriminant Analysis (LDA \cite{Runkler2012}), Support Vector Machines (SVM \cite{Chang2011}), and Binary Decision Trees (BDT \cite{Runkler2012}). The coding layer output of the AE and CAE are interpreted as a dimensionally reduced feature space and fed to the classifiers. Only the CNN is implemented as an end-to-end model and therefore already gives a classification as output.

The 36 implemented multidimensional \emph{hand-crafted features} comprise 212 dimensions that are extracted from the raw waveform data to form the feature space. All 212 dimensions are considered in creating the feature space for this model.

The \emph{AE} and \emph{CAE} model are used as a general automated way to reduce the dimensions of the high dimensional raw waveform data. Since a linear AE usually reaches the same performance as a principal component analysis (PCA) \cite{IanGoodfellow2016}, it is interesting to see whether the implemented multiple non-linear layers of the AE and CAE would result in any performance improvements compared to the PCA.
%\todo{write here the multiple layer architecture of which results exist for AE}

The \emph{CNN} model is fed with the raw waveform data to learn from different receptive field sizes. The training-set is again split into 80\,\% training samples and 20\,\% validation samples. This validation-set allows for a training performance monitoring after each training-epoch and is used to properly evaluate the current classification performance mid-training and enables training strategies such as early stopping and saving the best model.

\begin{figure}[htbp]%
\begin{minipage}{0.475\linewidth}%
\centering
\includegraphics[width=0.9\linewidth]{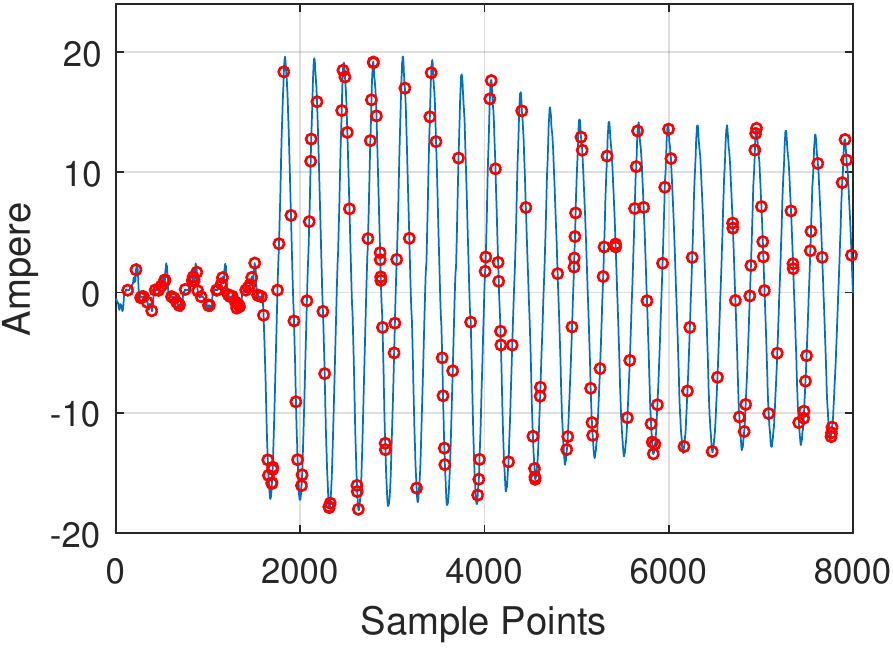}%
\end{minipage}
\hfill
\begin{minipage}{0.475\linewidth}%
\includegraphics[width=0.9\linewidth]{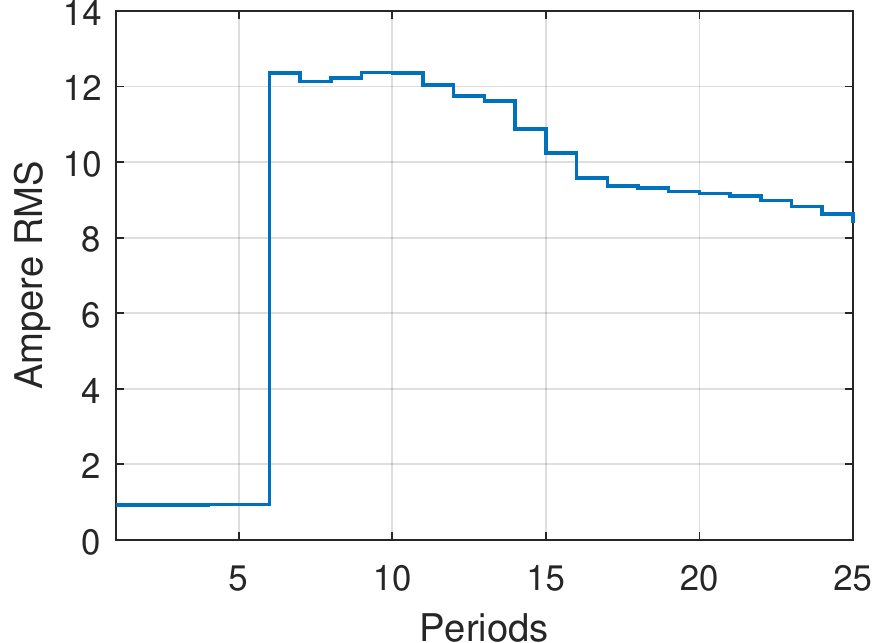}%
\end{minipage}%
\par
\medskip
\noindent
\begin{minipage}[t]{0.475\linewidth}
\caption{\textbf{Random sub-sampling:} The red circles show the randomly selected measurement points from the raw waveform current of a dishwasher event.}
\label{fig:randomSampling}
\end{minipage}
\hfill
\begin{minipage}[t]{0.475\linewidth}
\caption{\textbf{RMS-25:} The blue stepped line shows the current for each mains cycle in the measured 500\,ms segment of a dishwasher event.}
\label{fig:periods}
\end{minipage}%
\end{figure}

For the \emph{random sub-sampling} model, we extracted a random selected subset (without repetition) of the 500\,ms raw data samples (see Figure~\ref{fig:randomSampling}). To facilitate comparability to the hand-crafted feature set, the subset consists of 212 dimensions as well. This approach is equivalent to a non-equidistant sub-sampling and can be regarded as a very simple kind of feature extraction without any expert knowledge or comprehensive model architecture design.

Another model considers the root means square (RMS) energy of each mains cycle. The resulting 25 element long vector \emph{RMS-25} shows the actual absolute current over the 500\,ms (see Figure~\ref{fig:periods}). Since every appliance draws a different consumption during switch-on - the RMS of the 25 mains cycles is simple to calculate, but it is a powerful discriminative model in terms of appliance recognition \cite{Kahl2017}.

The \emph{principal component analysis} is a common method for reducing feature space dimensions. The PCA reduces a high dimensional data space into a lower one by descending ordered variances. These variances form a new cartesian coordinate system. To keep the comparability, the 212 highest variances are considered in this model. Assuming that both sets share the same distribution, the variances are calculated on the training-set and the resulting covariance matrix is used to transform the unseen test-set.

% \begin{figure}
% \centering
% \begin{minipage}{.475\linewidth}
% \centering
% \includegraphics[width=\textwidth]{randomSelectDims}
% \caption{\textbf{Random sub-sampling:} The red circles show the randomly selected measurement points from the raw waveform current of a dishwasher event.}
% \label{fig:randomSampling}
% \end{minipage}%
% \hfill
% \begin{minipage}{.475\linewidth}
% \centering
% \includegraphics[width=\textwidth]{periods}
% \caption{\textbf{RMS-25:} The blue stepped line shows the current for each mains cycle in the measured 500\,ms segment of a dishwasher event bla bla bla bla.}
% \label{fig:periods}
% \end{minipage}
% \end{figure}

% !TEX root = start_file.tex

\section{Results} \label{sec:Results}

\begin{figure}[htbp]
\centering
\includegraphics[width=0.85\linewidth]{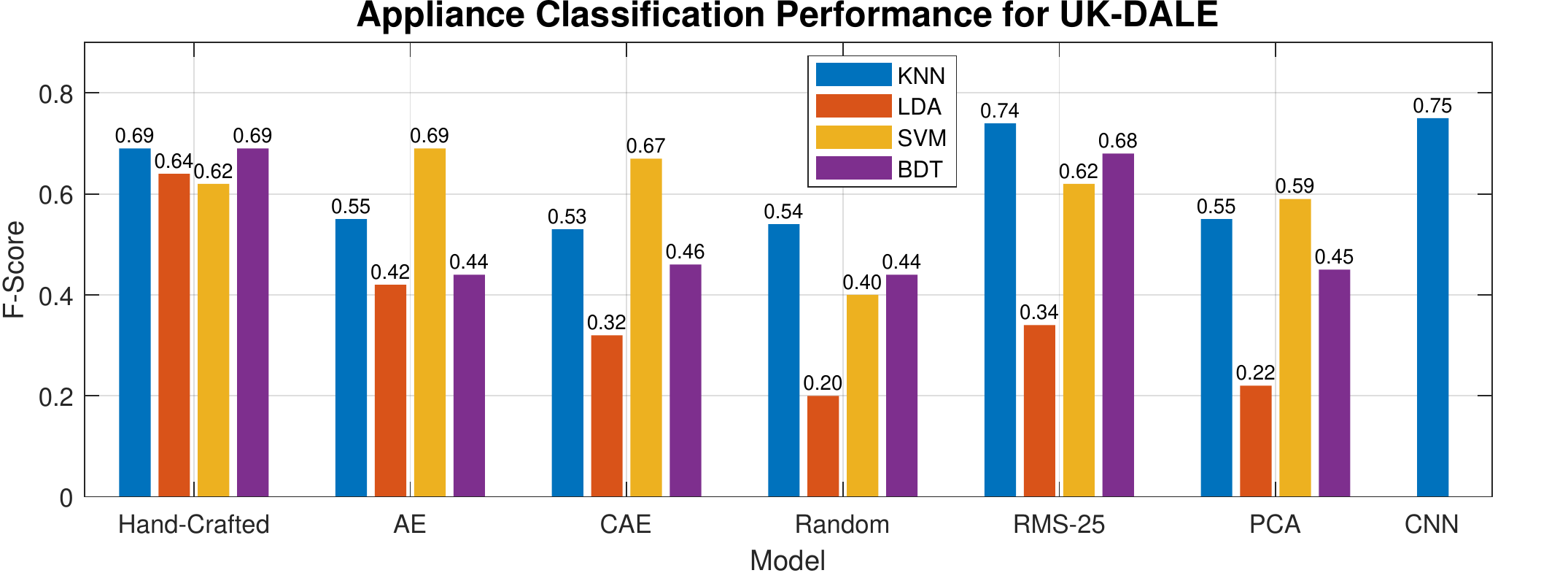}
\caption{Appliance classification performance using the seven introduced classification models and five classifiers on the UK-DALE dataset.}
\label{fig:barResultsUKDALE}
\end{figure}

\begin{figure}[htbp]
\centering
\includegraphics[width=0.85\linewidth]{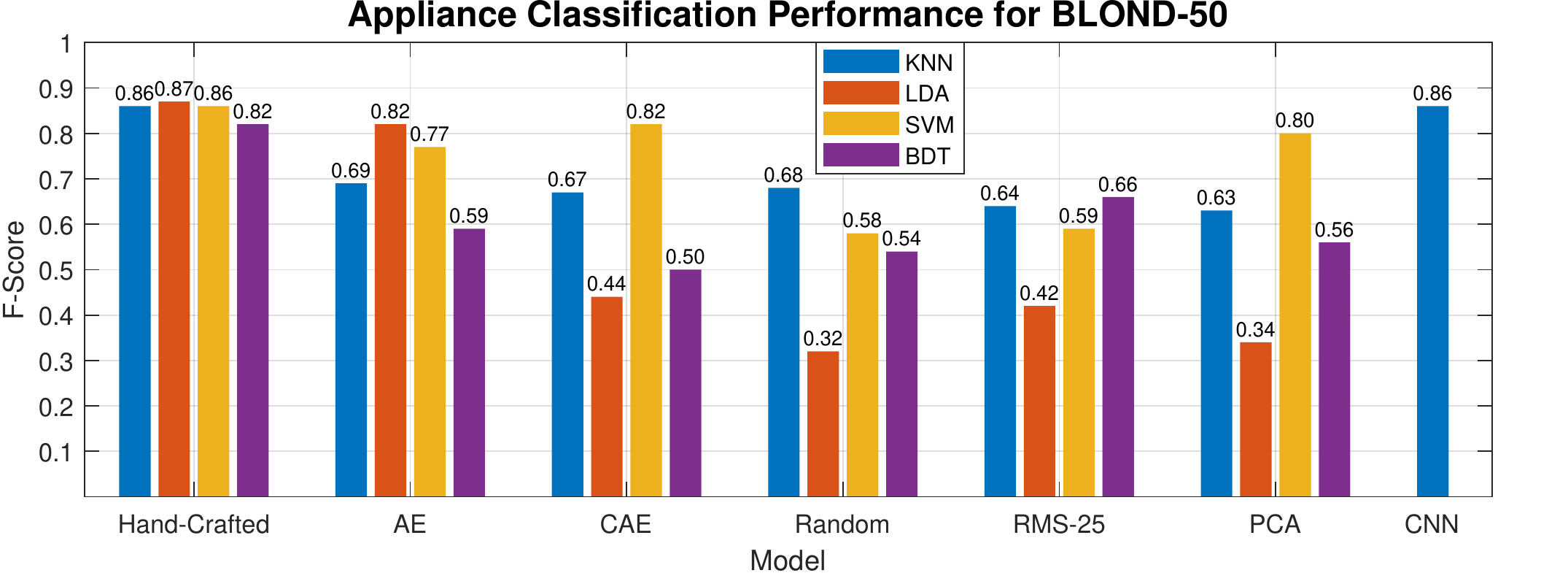}
\caption{Appliance classification performance using the seven introduced classification models and five classifiers on the BLOND-50 dataset.}
\label{fig:barResultsBLOND}
\end{figure}

The classification performance for all seven experiments is calculated using the predicted output of the corresponding model and its classifier. The results show a rather heterogeneous distribution of performance. The overall best performance including both datasets could be achieved with the CNN model. Regarding the stand-alone classifier, SVM and KNN reach the highest classification performance on average, with a mean F-Score over all six models with 0.60 for KNN on UK-DALE and 0.75 for SVM on BLOND-50 (see Table~\ref{tab:classifierresults}). The overall best classification performance could be achieved with 0.75 with the end-to-end CNN on UK-DALE and 0.87 with the hand-crafted features using the LDA classifier, closely followed by the CNN (see Figure~\ref{fig:barResultsUKDALE} and \ref{fig:barResultsBLOND}).

\vspace{-3mm}

Our interpretation of the results substantiates the observation: to replace the expert-driven hand-crafted feature extraction with a representation learning system, a large number of samples is necessary. Prevalent experiments on a lower number of appliance events led to much lower classification performance for the representation learning approaches. Humans are able to identify complex patterns and differences given only a few samples. The process of putting these patterns into metrics and numbers forms very powerful features, which is the main advantage of the expert-driven hand-crafted feature extraction.

%the data comes from table
\begin{table}[htbp]
\centering
\caption{Average classifier F-Score}
\label{tab:classifierresults}
\begin{tabular}{lcccc}
& KNN & LDA & SVM & BDT \\ \midrule
UK-DALE & 0.60 & 0.37 & 0.60 & 0.53 \\
BLOND-50 & 0.69 & 0.54 & 0.75 & 0.61 \\ \bottomrule
\end{tabular}
\end{table}

\begin{figure}[htbp]
\centering
\includegraphics[width=0.75\linewidth]{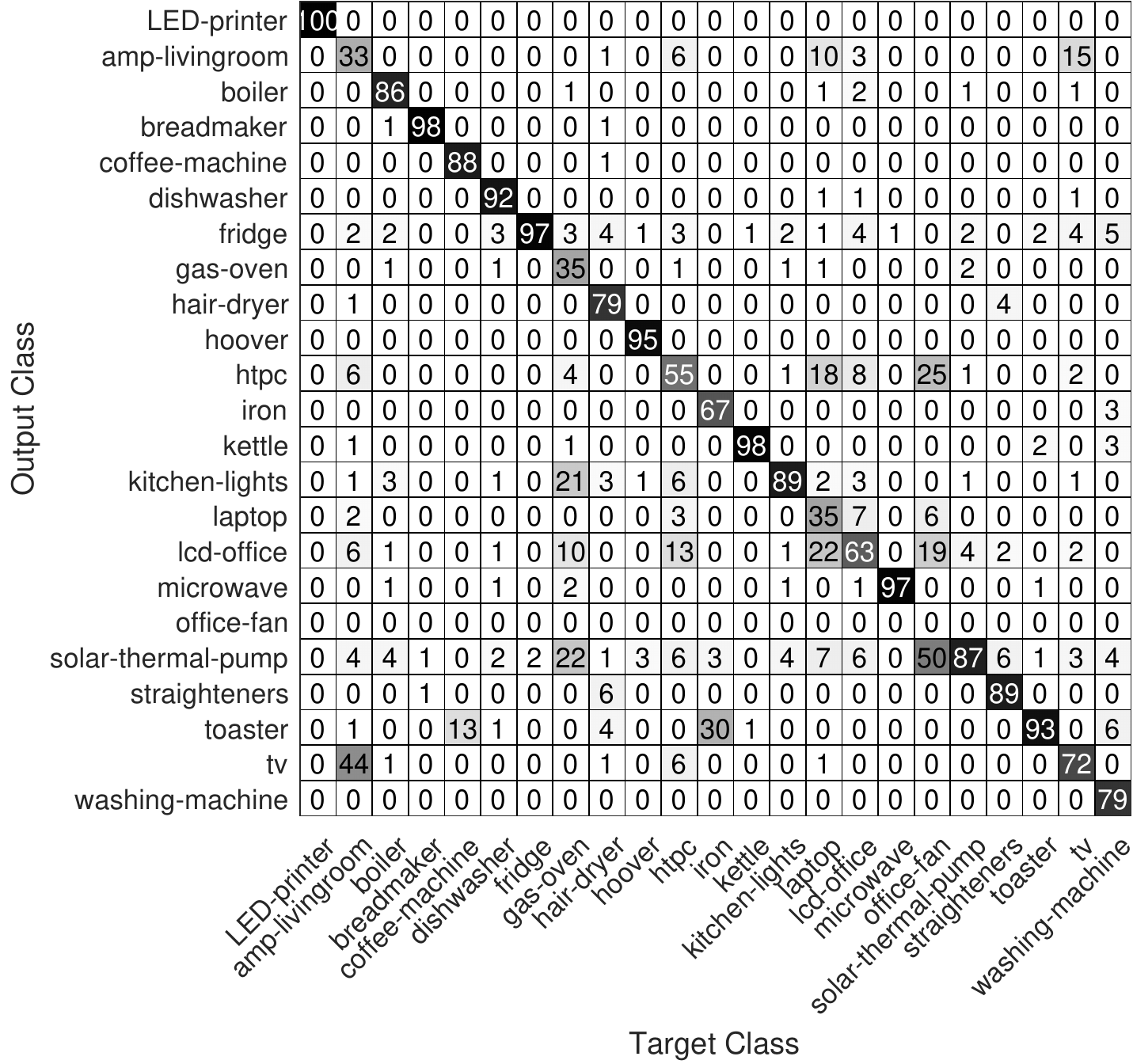}
\caption{The confusion matrix of the best performing CNN model (F-Score 0.75) shows the misclassification of each considered class of the UK-DALE dataset normalized to 100. Note that the values are rounded to integers}
\label{fig:confMatUKDALE}
\end{figure}

\begin{figure}[htbp]%
\begin{minipage}{0.28\linewidth}%
\centering
\includegraphics[width=0.75\linewidth]{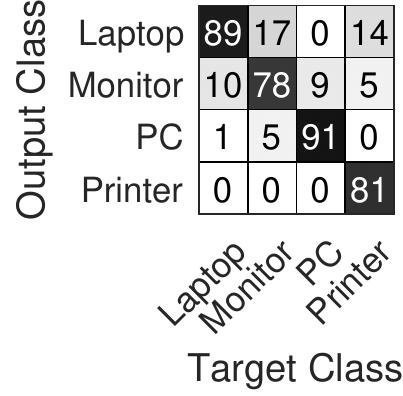}%
\end{minipage}
\quad
\begin{minipage}{0.67\linewidth}%
\centering
\includegraphics[width=0.9\linewidth]{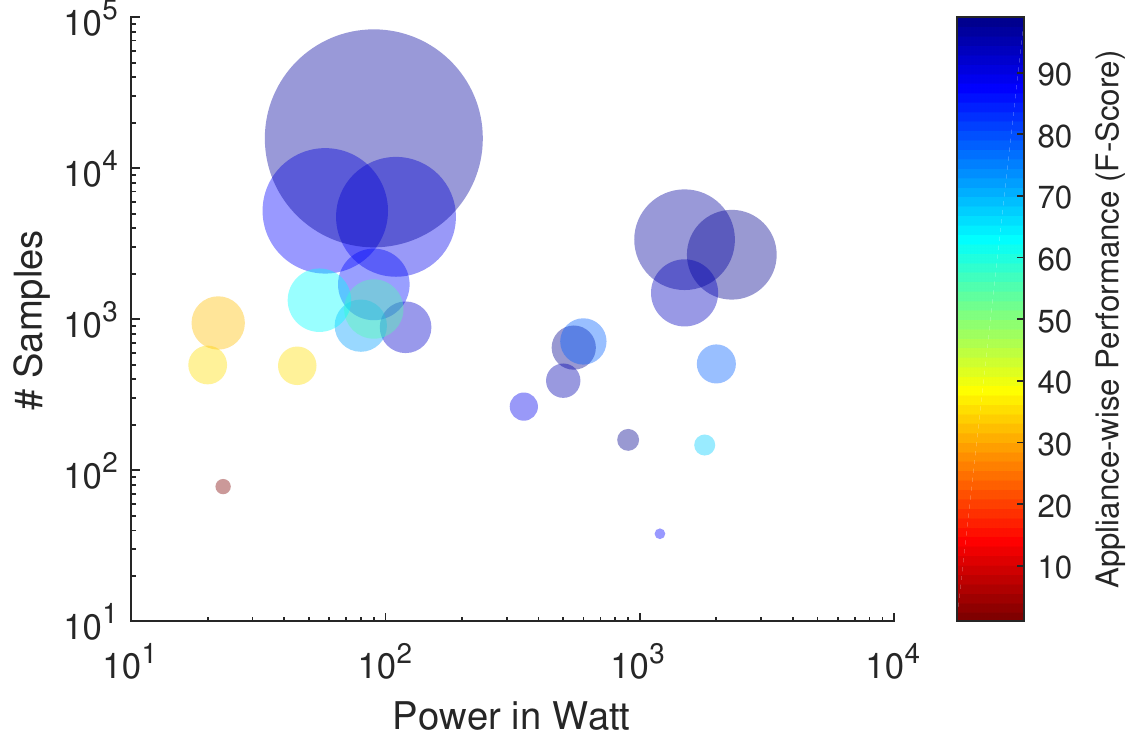}%
\end{minipage}%
\par
\medskip
\noindent
\begin{minipage}[t]{0.3\linewidth}
\caption{The confusion matrix of the best performing CNN model (F-Score 0.86) of BLOND-50, normalized to 100.}
\label{fig:CNNBLONDConfMat}
\end{minipage}
\hfill
\begin{minipage}[t]{0.65\linewidth}
\caption{The figure shows the dependencies of the sample count and appliance power to the actual recognition performance of the individual appliances from UK-DALE. The appliance-marker size correlates with the number of samples true to scale, showing the huge differences in the the frequency with which appliance events occur.}
\label{fig:dependencies}
\end{minipage}%
\end{figure}

% \begin{figure}[h]
% \centering
% \includegraphics[width=0.66\linewidth]{SamplePowerPerformance}
% \caption{The figure shows the dependencies of the sample number and appliance power to the actual recognition performance of the individual appliance.}
% \label{fig:confMatUKDALE}
% \end{figure}
%
% \begin{figure}[h]
% \centering
% \includegraphics[width=0.3\linewidth]{CNNBLONDConfMat}
% \caption{The confusion matrix of the best performing CNN model (F-Score 0.86) shows the misclassification of each considered class of the BLOND-50 dataset.}
% \label{fig:confMatBLOND}
% \end{figure}

\subsection{Classification Models}
Each classification model has been evaluated and performs differently for obvious and non-obvious reasons. %We applied further causal analysis on reasons for unexpected and significant results.

\begin{table*}[t]
\centering
\scriptsize
\caption{Neural network parameter of the best-performing architectures}
\label{tab:parameter}
\setlength{\tabcolsep}{1.3mm}
\begin{tabular}{lccccccccccc}
Dataset & Archit. & Dim-Scale per Layer & Batch Norm. & Activation & L2 Reg. & Norm. & Learn Rate & Batch Size & Noise & Optimizer & Loss Function \\ \midrule
UK-DALE & AE & {[}2,4,5 - 5,4,2{]} & yes & leaky relu & 0,00001 & variance & 0,0001 & 30 & 0,005 & ADAM \cite{Kingma2014} & MSE \\
BLOND-50 & AE & {[}10,5,2.5 - 2.5,5,10{]} & yes & leaky relu & 0,00001 & variance & 0,0001 & 45 & 0,005 & ADAM \cite{Kingma2014} & MSE \\ \midrule
UK-DALE & CAE & {[}5,4,2 - 2,4,5{]} & yes & leaky relu & - & variance & 0,001 & 45 & - & ADAM \cite{Kingma2014} & MSE \\
BLOND-50 & CAE & {[}5,5,5 - 5,5,5{]} & yes & leaky relu & - & variance & 0,001 & 45 & - & SGD & MSE \\ \midrule
UK-DALE & CNN & {[}5,2,2,2,2,2,2{]} & yes & leaky relu & - & variance & 0,001 & 30 & - & SGD & cat. cross-entr. \\
BLOND-50 & CNN & {[}5,5,5,2,2,2,2{]} & yes & leaky relu & - & variance & 0,001 & 30 & - & SGD & cat. cross-entr. \\ \bottomrule
\end{tabular}
\end{table*}

\subsubsection*{Hand-Crafted Features}
The best results with an F-Score of 0.69 could be achieved by using the max-normalization and the binary decision tree classifier. Figure~\ref{fig:barResultsUKDALE} and \ref{fig:barResultsBLOND} show a very homogeneous performance across the four classifiers, making the hand-crafted feature extraction a stable and the second best model in this benchmark.

\subsubsection*{AE}
The AE architecture with the best performance for UK-DALE comprises three encoding and decoding layers. The three fully connected encoding layers reduce the input by the factors 2, 4 and 5, similarly for the decoding layers. Since the performance with 0.69 for the best classifier is significantly higher compared to the PCA (0.59), some additional patterns in the feature space could be found by the non-linear layers. The best performance for BLOND-50 could be achieved with a three-layered architecture, with different reducing factors of 10, 5 and 2.5, similarly for the decoding layers and a batch size of 45 (see Table~\ref{tab:parameter} for further details).

\subsubsection*{CAE}
The expected performance improvement of the CAE due to its convolutional layers could not be reached in our experiments. We assume that the chosen parameter space was too far from the actual optimum. However, the best performing architecture and its parameters for the CAE in these experiments can be seen in Figure~\ref{fig:CAEArchitecture} and Table~\ref{tab:parameter}.

\subsubsection*{CNN}
The best performing end-to-end CNN architecture comprises the architecture of Figure~\ref{fig:ConvArchitecture} and the parameter settings of Table~\ref{tab:parameter}. The end-to-end implementation entails that the last layer of the neural network gives a classification as output. The fact that the parameter search gave the identical optimal parameter set for both datasets underlines a good generalization capability of the model.

\subsubsection*{Random Selected Raw Dimensions}
As expected and as the results show, this simple model of dimensional reduction does not allow a reliable and stable classification. With a mean F-Score of 0.39 for UK-DALE and 0.53 for BLOND-50, this model shows the worst performance in both cases. However, a pure random classification for the UK-DALE dataset would result in an F-Score of around 0.04, which is far below the performance of this model.

\subsubsection*{RMS-25}
The energy of the mains cycles forms a powerful feature that allows a very high classification performance in combination with a spectral metric \cite{Kahl2017}. Surprisingly, the KNN classifier using the mains cycles forms the second best model for the UK-DALE dataset. The appliances of the UK-DALE dataset can be well distinguished, based on their individual startup energy consumption pattern only. Unfortunately, in the case of the BLOND-50 dataset, the performance is only mid-range due to the different startup pattern of the individual appliances inside one appliance class.

\subsubsection*{PCA selected Dimensions}
PCA is one of the most applied methods for reducing the feature space \cite{Geron2017}. Therefore, the performance here is of interest. Since PCA is a linear transformation, not all information can be projected onto the lower feature space. Therefore, PCA usually performs worse than any well configured and trained neural network. Considering the simplicity of the algorithm and the absence of any expert knowledge, this still leaves PCA as an option.

\subsection{Appliances}
Regarding the best representation learning model (CNN) for UK-DALE, the average classification performance (F-Score) across all appliances lies at 0.75 (mean) and 0.86 (median). The four best recognized appliances are the \emph{kettle} (0.97), \emph{fridge} (0.97), \emph{microwave} (0.96) and \emph{breadmaker} (0.95). All of these appliances have in common that they are either represented by a huge number of samples or have a large power consumption.

The four worst recognized appliances are the \emph{gas-oven} (0.42), \emph{laptop} (0.41), \emph{amp-livingroom} (0.39) and \emph{office-fan} (0.0). Further analysis on these appliance events reveals that the \emph{amp-livingroom} shows one very short, small and heterogeneous peak transient while the \emph{gas-oven}, \emph{laptop} and \emph{office-fan} show a very low or even non-visible step in the power consumption. These observations and the fact that these four particular appliances have the lowest energy consumptions (see Figure~\ref{fig:dependencies}) of the whole appliance set, leads us to the assumption that their consumption is simply too low to distinguish properly from the background noise of the aggregated signal. The recognition of \emph{laptops} in BLOND-50 is significantly better, supporting the statement that the issue is regarded to these particular appliances. All the remaining appliances in UK-DALE were recognized correctly in most cases. Regarding BLOND-50, CNN could generalize very well over the multiple appliance models inside each class. The remaining misclassification of \emph{monitor} and \emph{laptop} are due to their similar power consumption.

%the amp has a very short and heterogeneous peak as startup-transient that is not sufficiently distinguishable from other smps caused transients
%the gas oven did not draw any visible transient characteristics in the aggregated signal - probably too low energy consumption
%the laptop did not draw any visible transient characteristics in the aggregated signal - probably too low energy consumption
%the fan did draw a barely visible power step in the aggregated signal - too low energy consumption

% !TEX root = start_file.tex

\section{Conclusions} \label{sec:Conclusions}
We presented an evaluation of several appliance classification models for two publicly available real-world energy consumption datasets. The classification models include conventional domain expert supported hand-crafted feature extraction, baseline-reference models and promising deep neural network models. The initial goal of the paper was to compare the classification performance of the classical machine learning approach and the more recent representation learning approaches.

The results of our experiments show comparable performance with a slight winning margin for the end-to-end implementation of the convolutional neural network (CNN). Our performance results support the statement that the representation learning approach is a worthy alternative to the classical machine learning processing-chain for appliance recognition systems in NILM. The effort for gaining expert-based features on the one side, neural network architecture and parameter search effort on the other side, as well as training data volume, are most likely the main decision criteria if the recognition system is based on a classical machine learning or a more recent representation learning framework.

% !TEX root = start_file.tex

\section*{Acknowledgements} \label{sec:Acknowledgements}
This research was partially funded by the Alexander von Humboldt Foundation established by the government of the Federal Republic of Germany and was supported by the Federal Ministry for Economic Affairs and Energy on the basis of a decision by the German Bundestag.

\FloatBarrier

%Bibliography
%\scriptsize
\bibliographystyle{unsrt}  
\bibliography{references}

\end{document}